\documentclass[aps]{revtex4} 
\usepackage{graphicx}

\begin{document}

\title{Bohr-Sommerfeld quantization and meson spectroscopy}
\author{F. Brau\thanks{E-mail: 
fabian.brau@umh.ac.be}}
\affiliation{Groupe de Physique Nucl\'eaire Th\'eorique, Universit\'e
de Mons-Hainaut, B-7000 Mons, Belgium}

\date{\today}

\begin{abstract}

We use the Bohr-Sommerfeld quantization approach in the context of constituent
quark models. This method provides, for the Cornell potential, analytical
formulae for the energy spectra which closely approximate numerical exact
calculations performed with the Schr\"{o}dinger or the spinless Salpeter
equations. The Bohr-Sommerfeld quantization procedure can also be used to
calculate other observables such as r.m.s. radius or wave function at the
origin. Asymptotic dependence of these observables on quantum numbers are also
obtained in the case of potentials which behave asymptotically as a power-law.
We discuss the constraints imposed by these formulae on the dynamics of the
quark-antiquark interaction.
\end{abstract}

\maketitle

\section{Introduction}
\label{sec:intro}

It is well known that nonrelativistic or semirelativistic constituent quark
models can describe with surprising accuracy a large part of the meson and
baryon properties. The mass spectra of hadrons have been intensively studied
since the pioneering works of, e.g., Eichten {\it et al.}~\cite{eich75},
Stanley and Robson~\cite{stan80} and Godfrey and Isgur~\cite{godf85} (see for
example~\cite{luch91,fulc94,brau98}). Other observables, such as decay widths,
have also been well
reproduced within the framework of potential models~\cite{godf85}. The success
of these calculations shows that it is possible to understand the major part
of the hadron observables using the simple picture for the quark-quark or
quark-antiquark interaction provided by constituent quark models.

In this paper, we will focus on the meson spectroscopy. Many potentials have
been proposed to describe the meson properties \cite{luch91,brau98}. Despite
of this diversity, the central part of the interaction used in usual models
presents similar characteristics: an attractive short range part with a
confining long range interaction. The prototype for these potentials is the
so-called Cornell potential~\cite{eich75} which can be used to describe a
large body of the meson masses except the masses of the pseudoscalar states
for which a spin dependent and a flavor mixing interaction is necessary.
Indeed, the role of this additional interaction is too important, in this
sector, to be neglected even in a first approximation. Recent lattice
calculations~\cite{bali97} confirm that the Cornell potential fits in a good
approximation the static quark-antiquark interaction.

Most of the models found in Refs.~\cite{luch91,brau98} have a phenomenological
nature: whereas the general
behavior of potentials are often dictated by physical considerations, the
values of their parameters are most of the time obtained by trial and error
from a comparison with some experimental data. The values of various
observables which can be generated from potential models will be strongly 
constrained by this general behavior of potentials and by the values of their
parameters; however, despite of the ease with which the numerical values of
these observables can be obtained, the exact nature of these constraints is
often very difficult to infer from the numerical results.

It is the purpose of this work to show that the use of a Bohr-Sommerfeld
quantization (BSQ) approach can provide unexpected insight into potential
models. We use this method to obtain some general characteristics of
constituent quark models. In particular, we will show that it is possible to 
obtain analytical expressions for the spectra, root mean square radii, decay
widths, electromagnetic mass splittings or electric polarizabilities, which
closely approximate the numerically exact results obtained from full quantum
Schr\"{o}dinger or spinless Salpeter approaches, and which can thus be used,
e.g., to derive accurate predictions for the dependence of the observables on
the quantum numbers. We will study the constraints imposed by these formulae
on the dynamics of the quark-antiquark interaction within a potential model.

\section{The BSQ Method}
\label{sec:theory}

The basic quantities in the BSQ approach \cite{tomo62} are the action
variables,
\begin{equation}
\label{eq1}
J_s=\oint p_s\, dq_s,
\end{equation}
where $s$ labels the degrees of freedom of the system, and where $q_s$ and
$p_s$ are the coordinates and conjugate momenta; the integral is performed
over one cycle of the motion. The action variables are quantized according to
the prescription
\begin{equation}
\label{eq2}
J_s=(n_s+c_s)\, h,
\end{equation}
where $h$ is Planck's constant, $n_s$ ($\geq 0$) is an integral quantum number
and $c_s$ is some real constant, which according to Langer should be taken
equal to $1/2$~\cite{lang37}.

To draw some general conclusions about potential models applied to meson 
spectroscopy we use the BSQ formalism, in the next sections, with the Cornell
potential. The properties of various observables are representative of those
which can be obtained from the usual constituent quark models. In particular,
conclusions about asymptotic behaviors (large value of quantum numbers) of
these quantities will be quite reliable since most of the potential models use
a linear confinement. Moreover, we will show that we can also obtain
information about asymptotic behaviors for any power-law confinement.

\subsection{Nonrelativistic calculations}
\label{subsec:firstf}

The nonrelativistic Hamiltonian corresponding to the Cornell potential reads
in natural units  
$(\hbar=c=1)$:
\begin{equation}
\label{eq4}
H=\frac{1}{2\mu}\left(p_r^2+\frac{p_{\phi}^2}{r^2}\right)-\frac{\kappa}{r}+a\,
r.
\end{equation} 
Of course since the interaction is central, the orbital angular momentum
$p_{\phi}=L$ is a constant of the motion. The expression for the radial
momentum $p_r$ derives from the conservation of the total energy $E$ of the
system,
\begin{equation}
\label{eq5}
p_{r}=\pm\, \frac{1}{r}\, \sqrt{-2\mu a\, r^3+2\mu E\, r^2 +2\mu \kappa\, r -
L^2}= \pm\, 
\frac{1}{r}\, \sqrt{g(r)}.
\end{equation}
The radial motion takes place between two turning points, $r_{-}$ and $r_{+}$,
which are the two positive zeros of $g(r)$; the three roots, $r_k$
($k=0,1,2$), of $g(r)$ are
\begin{equation}
\label{eq6}
r_k = 2\, \sqrt{\frac{-P}{3}}\, \cos\left(\frac{\theta + 2 \pi
k}{3}\right)+\frac{E}{3 a},
\end{equation}
where
\begin{eqnarray}
\label{eq6b}
\nonumber
P &=& -\frac{1}{3 a^2}\left(E^2+3a\kappa \right), \quad
\cos \theta = -\frac{Q}{2} \sqrt{\frac{27}{-P^3}},\\
Q &=& -\frac{1}{27 a^3}\left(2E^3 +9a\kappa E-27\frac{a^2}{2\mu}L^2 \right).
\end{eqnarray}
As $r_1 \leq 0 \leq r_2 \leq r_0$, the turning points are $r_{-}=r_2$ and
$r_{+}=r_0$. 

Quantization of $J_{\phi}$ trivially gives $L=\ell+c_{\phi}$; on the other
hand, quantization of $J_r$ leads to the equation
\begin{eqnarray}
\label{eq7}
\nonumber
2\mu\, r_{+}\, (E r_1+2\kappa)\, K(\eta)+2\mu\, E\, r_{+}\,(r_{+}-r_{1})\,
E(\eta)-3
(\ell+c_{\phi})^2\, \Pi(\pi/2,\gamma,\eta)- \\
\frac{3}{2}\pi\, (n+c_{r}) r_{+}\sqrt{r_{+}-r_{1}}\, \sqrt{2\mu\, a}=0,
\end{eqnarray}
where
\begin{equation}
\label{eq8}
\eta=\frac{r_{+}-r_{-}}{r_{+}-r_{1}} \quad \text{and} \quad
\gamma=1-\frac{r_{-}}{r_{+}},
\end{equation}
$K(x)$, $E(x)$ and $\Pi(\pi/2,x,y)$ are the complete elliptic integrals of the
first, second and third kind, respectively \cite [p. 904]{grad80}, and $n$ is
the radial quantum number. This appears to be a rather complicated equation
since it cannot be solved explicitly for the energy. However, it leads to very
accurate results if we choose the Langer prescription $c_r=c_{\phi}=1/2$, as
it can be seen in Table~\ref{tab:1} where a selected set of masses, computed
from Eq.~(\ref{eq7}), are compared with the results of an exact calculation 
\cite{fulc94}. The accuracy of the results presented in Table~\ref{tab:1} is
representative of that obtained for the states which are not displayed here.
One sees that, even for small quantum numbers, the errors introduced by the
BSQ approximation are remarkably small; indeed the errors in the spectrum
reported in Table~\ref{tab:1} do not exceed $0.1\%$.

One of the most useful byproducts of the BSQ method is to provide simple
asymptotic expressions for the total energy for large values of $\ell$
and $n$. Indeed, for large angular momenta ($\ell \gg n$), the orbits become
almost circular, and thus $r_{-} \approx r_{+}$ and $\eta \approx 0$. In this
limit, $\theta = \pi$ and all the elliptic integrals are equal to $\pi /2$;
keeping the leading terms in Eq.~(\ref{eq7}), we obtain:
\begin{equation}
\label{eq9}
E \sim \frac{3}{2} \left(\frac{a^2}{\mu}\right)^{1/3} \ell^{2/3} \quad \quad
(\ell \gg n).
\end{equation}
This is a well-known result: a linear potential does not lead to linear Regge
trajectories in a nonrelativistic description; correct Regge trajectories can
only be obtained with confining potentials rising as $r^{2/3}$ \cite{fabr88}.

For large radial quantum numbers ($n \gg \ell$), the orbits have a large
eccentricity, and thus $r_1 \approx r_{-} \approx 0$ and $\eta \approx 1$.
Although $\eta=1$ cannot be directly inserted into Eq.~(\ref{eq7}) because of
the singularity of the elliptic integrals, setting $r_1 = r_{-} = 0$ into
Eq.~(\ref{eq1}) and evaluating an elementary integral leads to:
\begin{equation}
\label{eq10}
E \sim \left(\frac{3\pi}{2}\right)^{2/3} \left(\frac{a^2}{2\mu}
\right)^{1/3} n^{2/3} \quad \quad (n \gg \ell).
\end{equation}
This is a new result: the radial quantum number does not appear explicitly in
the classical Hamiltonian, and thus ``naive" semiclassical methods fail to
reproduce correctly the spectrum in this sector \cite{shee81,olso97}.
Eqs.~(\ref{eq9}) and~(\ref{eq10}) show that the spectrum has the same
asymptotic behavior in $\ell$ \textit{and} $n$. The square of the ratio of the
``slopes", $R$, assumes the value:
\begin{equation}
\label{eq10b}
R= \left(\frac{\pi^2}{3}\right)^{2/3},
\end{equation}
that is, it does not depend on any physical parameter. In fact, these
properties remain valid for any power-law confining potential. Indeed for
$V(r)\sim ar^{\alpha}$ ($\alpha > 0$), the large-$\ell$ behavior is found to
be:
\begin{equation}
\label{eq11}
E \sim \left(\frac{a^2}{(2\mu)^{\alpha}}\right)^{1/(\alpha+2)} 
\left(1+\frac{\alpha}{2}\right)
\left(\frac{2}{\alpha}\right)^{\alpha/(\alpha+2)}  
 \ell^{2\alpha/(\alpha+2)} \quad (\ell \gg n),
\end{equation}
confirming the result of Ref.~\cite{fabr88}; for large $n$, the turning points
behave as:
\begin{eqnarray}
\label{eq11b}
\nonumber
r_{+} &\sim& \left(\frac{E}{a}\right)^{1/\alpha}, \\
r_{-} &\sim& 0,
\end{eqnarray}
and Eq.~(\ref{eq1}) leads to an elementary integral which gives:
\begin{equation}
\label{eq12}
E \sim \left(\frac{a^2}{(2\mu)^{\alpha}}\right)^{1/(\alpha+2)} 
\left(\frac{\alpha \pi}{B(1/\alpha,3/2)}\ n\right)^{2\alpha/(\alpha+2)}\quad
(n \gg \ell),
\end{equation}
where $B(x,y)$ denotes the beta function \cite [p. 948]{grad80}. We will
discuss in Sec.~\ref{subsec:thirdf} the implication of these results for the
meson spectroscopy.

The BSQ method can be used to compute other observables as well. In quantum
theory, the value of an observable $A$ is obtained from the wave function of
the system as an average value. In the BSQ context, we have in contrast to
evaluate an average over time according to
\begin{equation}
\label{eq17}
\langle A \rangle = \frac{1}{T} \int_0^{T}\, A(t)\, dt,
\end{equation}
where $T$ is the period of the radial motion. As an example, we calculate the
mean square radii of the states bound by the Cornell potential, which were
reported in Table \ref{tab:1}.
Using the nonrelativistic equation of motion $\dot{r}=p_r/\mu$, we can write
\begin{equation}
\label{eq18}
\langle r^2 \rangle = \frac{2\mu}{T} \int_{r_{-}}^{r_{+}}\, \frac{r^2}{p_r}\,
dr,
\end{equation}
and
\begin{equation}
\label{eq19}
T = 2\mu \int_{r_{-}}^{r_{+}}\, \frac{1}{p_r}\, dr.
\end{equation}
One obtains:
\begin{equation}
\label{eq20}
\langle r^2 \rangle=\frac{1}{15 a^2}\left[9 a\kappa+8 E^2+\left(4
E\kappa-\frac{3 a 
(\ell+c_{\phi})^2}{\mu}\right)\frac{K(\eta)}{r_{1}K(\eta)+(r_{+}-r_{1})E(\eta)
}\right].
\end{equation}
Inspection of Table \ref{tab:1} shows again that the accuracy of this formula
is excellent. It can be used to derive nice asymptotic formulae; indeed one
obtains
\begin{equation}
\label{eq20b}
\langle r^2 \rangle\sim \left(\frac{1}{a \mu}\right)^{2/3} \, \ell^{4/3}\quad
(\ell \gg n),
\end{equation}
\begin{equation}
\label{eq20c}
\langle r^2 \rangle\sim \left(\frac{1}{a \mu}\right)^{2/3} \, 
\frac{2}{5}\left(3\pi^4\right)^{1/3}\, n^{4/3}\quad (n \gg \ell).
\end{equation}
The behavior of the r.m.s. radius is the same for large $\ell$ {\it and} $n$.
Comparison with Eqs.~(\ref{eq9}-\ref{eq10}) shows that the r.m.s. radius
becomes proportional at large $\ell$ or $n$ to the total energy. This behavior
is a property of a linear potential. The ratio of the ``slopes" of the
relations~(\ref{eq20b}-\ref{eq20c}) does not depend on any physical parameter
as it was the case for the ratio $R$ for the energy. These properties are
still valid for any power-law confining potential. For $V(r)\sim ar^{\alpha}$, 
the large-$\ell$ and large-$n$ asymptotic formulae are:
\begin{equation}
\label{eq20d}
\langle r^2 \rangle\sim \left(\frac{1}{\alpha a \mu}\right)^{2/(\alpha+2)} \,
\ell^{4/(\alpha+2)}\quad (\ell \gg n),
\end{equation}
\begin{equation}
\label{eq20e}
\langle r^2 \rangle\sim \left(\frac{1}{2 a \mu}\right)^{2/(\alpha+2)} \,
\frac{B(3/\alpha,1/2)}{B(1/\alpha,1/2)}\,
\left(\frac{\alpha \pi}{B(1/\alpha,3/2)}\ n\right)^{4/(\alpha+2)} \quad (n \gg
\ell).
\end{equation}
These expressions can also be used to study the electric polarizability of
mesons since the latter is proportional to $\langle r^2
\rangle^2$~\cite{luch91}.

Other observables can also be evaluated like, e.g., the various decay widths
of the system. In the nonrelativistic reduction, the main ingredient in the
calculation of these quantities is the modulus of the $\ell=0$ wave function
at the origin, $|\Psi(0)|^2$, a quantity which is not available in our 
formalism. However, use of the following relation \cite{luch91}, valid in the
nonrelativistic case
\begin{equation}
\label{eq21}
|\Psi(0)|^2=\frac{\mu}{2\pi} \left\langle \frac{dV(r)}{dr} \right\rangle,
\end{equation}
makes possible an analytical evaluation of the decay widths within the BSQ
approximation. For a Cornell potential, using the definition (\ref{eq17}) and
the equation of motion $\dot{r}=p_r/\mu$, the square of the modulus of the
wave function at the origin is found to be
\begin{equation}
\label{eq21b}
|\Psi(0)|^2=\frac{\mu}{2\pi} \left(a+\kappa \left\langle 1/r^2
\right\rangle\right),
\end{equation}
with
\begin{equation}
\label{eq21c}
\left\langle \frac{1}{r^2}
\right\rangle=\frac{\Pi(\pi/2,\gamma,\eta)}{r_{+}(r_1\,
K(\eta)+(r_{+}-r_1) E(\eta))}, 
\end{equation}
where $\gamma$ and $\eta$ are defined by Eq.~(\ref{eq8}). In
Table~\ref{tab:1}, we show a comparison between this formula and values
obtained with an exact calculation using parameters of Ref.~\cite{fulc94}. The
accuracy is here also remarkable especially for light mesons, with an error
smaller than $1\%$ for the $n\bar{n}$ states. The error is about $5\%$ for the
$b\bar{b}$ ground state. In this sector the Coulomb part of the interaction
plays a more active
role and more important errors are introduced since the BSQ result is only
exact for a pure linear potential. However, the error for the ratio
$|\Psi_{1S}(0)/\Psi_{2S}(0)|^2$ is always smaller than $1\%$ even for the
heavy mesons.

It is also possible to derive an asymptotic behavior formula for $|\Psi(0)|^2$
for a potential which behaves as $V(r) \sim ar^{\alpha}$:
\begin{equation}
\label{eq21d}
|\Psi(0)|^2\sim \frac{\mu\alpha\, a^{1/\alpha}}{\pi B(1/\alpha,1/2)}\
E^{(\alpha-1)/\alpha} \quad \quad (n \gg \ell), 
\end{equation}
with the total energy $E$ given by Eq.~(\ref{eq12}). The asymptotic value for
the ratio of the wave function at the origin, for two states with radial
quantum numbers equal to $m$ and $n$, reads
\begin{equation}
\label{eq21e}
\left|\frac{\Psi_m(0)}{\Psi_n(0)}\right|^2\sim
\left(\frac{m}{n}\right)^{2(\alpha-
1)/(\alpha+2)}.
\end{equation}
It depends only on the value of $\alpha$ and the radial quantum
numbers considered.

Note that the BSQ approach can also provide analytical formulae for electric
mass splittings because the wave function at the origin and the mean value of
$1/r$, which can be calculated with Eq.~(\ref{eq17}), are the main ingredients
for the evaluation of this quantity.

\subsection{Semirelativistic calculations}
\label{subsec:secondf}

Similar calculations can be performed within a relativistic kinematics.
The semirelativistic Hamiltonian corresponding to the nonrelativistic
Hamiltonian (\ref{eq4}) reads
\begin{equation}
\label{eq22}
H=2\sqrt{p_r^2+\frac{p_{\phi}^2}{r^2}+m^2}-\frac{\kappa}{r}+a\, r.
\end{equation}
The orbital angular momentum $p_{\phi}=L$ is still a constant of the motion.
The expression of the radial momentum $p_r$ derives from the conservation of
the total energy $M$ of the system,
\begin{equation}
\label{eq23}
p_{r}=\pm\, \frac{1}{2r}\, \sqrt{a^2 r^4-2 a\, M r^3+\left(M^2-2a\kappa-4
m^2\right)\, r^2 +2M \kappa\, r 
+\kappa^2- 4 L^2}\equiv \pm\, \frac{1}{2r}\, \sqrt{h(r)},
\end{equation}
The polynomial $h(r)$ is here of the fourth order. One can verify that $h(r)$
reduces to $g(r)$ in the nonrelativistic limit ($m \rightarrow \infty$). The
radial motion takes place between two turning points, $r_{-}$ and $r_{+}$,
which are two zeros of $h(r)$; the four roots of $h(r)$ are:
\begin{eqnarray}
\label{eq23b}
\nonumber
r_1 &=& -\frac{1}{2}\left(\sqrt{U}+\sqrt{\Delta_{+}}\right)+\frac{M}{2 a},
\quad
r_2 = -\frac{1}{2}\left(\sqrt{U}-\sqrt{\Delta_{+}}\right)+\frac{M}{2 a},\\
r_3 &=& \frac{1}{2}\left(\sqrt{U}-\sqrt{\Delta_{-}}\right)+\frac{M}{2 a},
\quad
r_4 = \frac{1}{2}\left(\sqrt{U}+\sqrt{\Delta_{-}}\right)+\frac{M}{2 a},
\end{eqnarray}
with,
\begin{eqnarray}
\label{eq23c}
\nonumber
\Delta_{\pm}&=&-(U+2P)\pm \frac{2Q}{\sqrt{U}}\\
\nonumber
P &=& -\frac{1}{2 a^2}\left(M^2+4a\kappa+8m^2 \right), \quad
Q = -\frac{4m^2 M}{a^3}\\
\nonumber
U &=& 2 \sqrt{\frac{-S}{3}} \cos\left(\frac{\theta}{3}\right)-\frac{2P}{3},
\quad
\cos\theta = -\frac{T}{2} \sqrt{\frac{27}{-S^3}}\\
\nonumber
S &=& -\frac{P^2}{3}-4R, \quad
T = -\frac{2}{27}P^3+\frac{8}{3}P R-Q^2\\
R &=& \frac{1}{16 a^4}\left(M^4+8M^2\left(a\kappa-2m^2 \right)+16a^2
\left(\kappa^2-4L^2\right)\right).
\end{eqnarray}
The two turning points are $r_{-}=r_2$ and $r_{+}=r_3$.

Quantization of $J_{\phi}$ leads to $L=\ell+c_{\phi}$; on the other hand
quantization of $J_r$ gives the equation
\begin{eqnarray}
\label{eq24}
\nonumber
\alpha_1\, K(\eta)+\alpha_2\, \Pi(\pi/2,\gamma,\eta)+\alpha_3\,
\Pi\left(\pi/2,\frac{r_1}{r_{-}}
\gamma,\eta\right)+\alpha_4\, E(\eta)\\ 
-2\pi a(n+c_{r})\,\sqrt{r_4-r_{-}} \sqrt{r_{+}-r_1}=0,  
\end{eqnarray}
where
\begin{equation}
\label{eq25}
\eta=\frac{r_{4}-r_{1}}{r_{4}-r_{-}}\, \frac{r_{+}-r_{-}}{r_{+}-r_{1}}, \quad 
\gamma=\frac{r_{+}-r_{-}}{r_{+}-r_1},
\end{equation}
and where
\begin{eqnarray}
\label{eq26}
\nonumber
\alpha_1 &=& \frac{a}{2}(r_{+}-r_1)(r_{-}-r_1)(M+a(r_4-r_1)),\\
\nonumber
\alpha_2 &=& -2(r_{-}-r_1)\left(a\kappa+2m^2\right),\\
\nonumber
\alpha_3 &=& -2a^2 r_{+} r_4 (r_{-}-r_1),\\
\alpha_4 &=& \frac{aM}{2} (r_4-r_{-})(r_{+}-r_1).
\end{eqnarray}

This equation present some common features with the non-relativistic formula
(\ref{eq7}): it involves complete elliptic integrals, it cannot be solved
explicitly for the energy and it leads to very accurate results if we choose
the Langer prescription $c_{r}=c_{\phi}=1/2$. A comparison of the results
obtained from this equation and exact calculations \cite{fulc94} is given in
Table~\ref{tab:2}. The accuracy is less good for the ground states in
the relativistic case. The poorest result is obtained for the $\rho$ state
with an error of about $2.8\%$. But the convergence is quite rapid since the
error already reduces to about $0.2\%$ for the $\rho(1450)$ meson. 

We can also obtain some simple asymptotic expressions for the masses of the
states for large values of $\ell$ and $n$. The condition for circular orbits,
$r_{-}=r_{+}$, leads to
\begin{equation}
\label{eq27}
2\sqrt{U}=\sqrt{\Delta_{+}}+\sqrt{\Delta_{-}}.
\end{equation}
Since the values of the total energy $M$ and the angular momentum $\ell$ are
important, we have
\begin{equation}
\label{eq28}
\Delta_{+}=\Delta_{-}=-(U+2P).
\end{equation}
These two last equations give $U=-P$ from which we obtain an expression for
$\cos (\theta/3)$. Comparison of this expression with the definition of $\cos
\theta$ (\ref{eq23c}) imposes $R=0$, and we find
\begin{equation}
\label{eq29}
M\sim 2\sqrt{2} \sqrt{a\, \ell} \quad (\ell \gg n).
\end{equation}
This is the expected result \cite{luch91,kang75,mart86}: a linear potential
leads to linear Regge trajectories in the relativistic case.

For large values of the radial quantum number, the large eccentricity of the
orbits implies $r_1 \sim r_{-} \sim 0$ and $r_{+} \sim r_{4} \sim M/a$.
Setting $r_1 = r_{-} = 0$ into Eq.~(\ref{eq1}) and evaluating an elementary
integral leads to:
\begin{equation}
\label{eq30}
M\sim 2\sqrt{\pi} \sqrt{a\, n} \quad (n \gg \ell).
\end{equation}
This result shows that the spectrum has the same asymptotic behavior in $\ell$
{\it and} $n$. The square of the ratio of the slopes $R$ assumes the value
\begin{equation}
\label{eq31}
R=\pi/2,
\end{equation}
that is, this quantity is still parameter-independent. In fact, as in the
nonrelativistic case, these properties remain valid for any power-law
confining potential. For $V(r)\sim ar^{\alpha} (\alpha > 0)$, the large-$\ell$
behavior is found to be
\begin{equation}
\label{eq32}
M \sim  a^{1/(\alpha+1)}
(\alpha+1)\left(\frac{2\ell}{\alpha}\right)^{\alpha/(\alpha+1)} \quad (\ell
\gg n);
\end{equation}
for large $n$, the turning points have the forms
\begin{eqnarray}
\label{eq33}
\nonumber
r_{+} &\sim& \left(\frac{M}{a}\right)^{1/\alpha}, \\
r_{-} &\sim& 0,
\end{eqnarray}
and Eq.~(\ref{eq1}) leads to an elementary integral which gives:
\begin{equation}
\label{eq33b}
M \sim a^{1/(\alpha+1)} \left(\frac{2\pi(\alpha+1)}{\alpha}
n\right)^{\alpha/(\alpha+1)} \quad (n \gg \ell).
\end{equation}

To calculate the r.m.s. radius in the semirelativistic formulation we use the
definition (\ref{eq17}) with the equation of motion $\dot{r}=4 p_r/(M-V(r))$.
The relativistic version of Eq.~(\ref{eq18}) is found to be
\begin{equation}
\label{eq34}
\langle r^2 \rangle=\frac{1}{2T}\int_{r_{-}}^{r_{+}}\, \frac{r^2}{p_r}\,
(M-V(r))\, dr,
\end{equation}
with
\begin{equation}
\label{eq35}
T=\frac{1}{2}\int_{r_{-}}^{r_{+}}\, \frac{(M-V(r))}{p_r}\, dr.
\end{equation}
Using the expression~(\ref{eq23}) of the radial momentum $p_r$ the
integration of Eqs.~(\ref{eq34}-\ref{eq35}) leads to
\begin{equation}
\label{eq36}
\langle r^2 \rangle =\frac{1}{3a^2}\, \frac{\beta_1 K(\eta)+\beta_2
\Pi\left(\pi/2,
\gamma,\eta\right)+\beta_3E(\eta)}{\beta_4 K(\eta)+\beta_5 E(\eta)},
\end{equation}
with
\begin{eqnarray}
\label{eq37}
\nonumber
\beta_1 &=& - 2Mr_1\left(M^2+a\kappa-4m^2\right)-2M^2\kappa-2a\left(\kappa^2-
4(\ell+c_{\phi})^2\right)\\
\nonumber
&+& a\left(M^2+a\kappa+8m^2\right)(r_1(r_{+}+r_1)-r_{-}(r_{+}-r_1)),\\
\nonumber
\beta_2 &=& 24M m^2(r_{-}-r_1),\\
\nonumber
\beta_3 &=& -a\left(M^2+a\kappa+8m^2\right)(r_4-r_{-})(r_{+}-r_1),\\
\nonumber
\beta_4 &=& -2M r_1-2\kappa +a(r_1(r_{+}+r_1)-r_{-}(r_{+}-r_1)),\\
\beta_5 &=& -a(r_4-r_{-})(r_{+}-r_1).
\end{eqnarray}
This rather complicated equation gives very accurate results as it can be seen
in Table~\ref{tab:2} where they are compared with exact calculations. It can
be used to obtain some simple asymptotic formulae:
\begin{equation}
\label{eq38}
\langle r^2 \rangle\sim \frac{2}{a} \, \ell \quad (\ell \gg n),
\end{equation}
\begin{equation}
\label{eq39}
\langle r^2 \rangle\sim \frac{4\pi}{3 a}\, n \quad (n \gg \ell).
\end{equation}
Like in the nonrelativistic calculations the asymptotic behavior is the same
in $\ell$ {\it and} $n$. Comparison with Eqs.~(\ref{eq29}-\ref{eq30}) shows
that the r.m.s. radius becomes proportional to the total energy for large
values of quantum numbers, which is a characteristic of a linear potential.
The ratio of the slopes which appear in these asymptotic formulae is
parameter-independent. These properties are, here also, still valid for a
power-law potential $V(r) \sim ar^{\alpha}$. In this case the asymptotic
formulae are
\begin{equation}
\label{eq40}
\langle r^2 \rangle\sim \left(\frac{2}{a \alpha}\right)^{2/(\alpha+1)} \,
\ell^{2/(\alpha+1)} \quad (\ell \gg n),
\end{equation}
\begin{equation}
\label{eq41}
\langle r^2 \rangle\sim \frac{1}{3} \left(\frac{2\pi (\alpha+1)}{a \alpha}\,
n\right)^{2/(\alpha+1)}\quad (n \gg \ell).
\end{equation}

\subsection{Discussion}
\label{subsec:thirdf}

Application of the results obtained in Sec.~\ref{subsec:firstf} and
\ref{subsec:secondf} to meson spectroscopy could prove to be of a great
interest. Indeed, if we want to produce linear Regge trajectories (which are
well observed experimentally in the light meson sector) within a potential
model, we must use a potential which behaves at large $r$ like $r^{2/3}$ in
the nonrelativistic case, and like $r$ for a semirelativistic kinematics. In
both cases, we have just shown that the trajectories for radial excitation are
then necessarily also linear; moreover, the square of the ratio of the slopes
$R$ is completely determined by the asymptotic behavior of the potential and
the kinematics used, and takes here the value 
\begin{equation}
\label{eq42}
R=\sqrt{3}
\end{equation}
in the nonrelativistic case and
\begin{equation}
\label{eq43}
R=\pi/2
\end{equation}
in the semirelativistic case. This is an additional strong constraint,
especially as these ratios are parameter-independent. But, since we have
chosen the confining potential to reproduce the energy orbital trajectories 
(Regge trajectories), the asymptotic behavior of orbital and radial
trajectories of other observables is determined. These results imply that the
currently investigated potential models~\cite{luch91,brau98}, which use
power-law confining interactions, can only describe a restricted class of
experimental data. This remark could prove decisive in the (near?) future,
when new experimental informations on the radial excitations of light mesons,
which are still very scarce, become available. For example, if the
experimental energy radial trajectories differ from a straight line, or if $R$
does not assume the values of Eqs.~(\ref{eq42}) or (\ref{eq43}), the
understanding of the physics underlying the confinement could become more
problematic, and a simple power-law potential would not be sufficient to
describe the confinement of quarks (the above arguments remain valid if one
considers the possibility that the asymptotic behavior of the Regge trajectory
could not be exactly linear). 

If the experimental radial trajectories prove in the end to be linear, but if
the ratio $R$ differs significantly from the values of Eqs.~(\ref{eq42}) or
(\ref{eq43}), the introduction of a scalar component in the confining
potential could be a possible way out; indeed we have performed calculations
using the BSQ method, which show that this additional flexibility (which only
makes sense within a relativistic approach) allows the ratio $R$ to take any
value between $\pi/2$ and $2$. Of course, a quantum number-dependent confining
potential could easily lead to a complete decoupling of the Regge and radial
trajectories. A better kinematical treatment of the problem, as the use of
full covariant equations (see for example Ref.~\cite{salp51}), could also
alter the predicted large-$\ell$ and large-$n$ behaviors of the trajectories.  

A limited set of additional experimental informations on the radial
excitations of light mesons could already be sufficient to draw important
conclusions. Indeed, the linear behavior of the Regge trajectories, which is
governed by the long range part of the interaction, is already reached
experimentally for the very first orbital excitations. The situation seems to
be even more favorable for the radial trajectories, since, as the masses grow
faster as a function of $n$ than as a function of $\ell$, the asymptotic
regime is expected to be attained at very low $n$; most potential
models~\cite{luch91,brau98} predict that this regime is already effective from
the first radial excitation.

To conclude this discussion, it is worth noting that the calculations
presented in this paper allow to understand some well-observed properties of
potential models. It is known that the use of a semirelativistic kinematics
yields a better description of radial excitations such as
$\rho(1450)$ for the mesons and $N(1440)$ for the baryons. Actually a
non-relativistic description gives in general too high masses for these states
(see for example~\cite{ono82,blas90,sema97}). Indeed, correct description of
Regge trajectories leads to heavier masses for radial excitations
in nonrelativistic calculations since Eqs.~(\ref{eq42}) and (\ref{eq43}) show
that the ratio $R$ is higher when the Schr\"{o}dinger equation is used.

It is also known that experimentally the mass of the $K^*(892)$ meson is in
good approximation the average of the masses of the $\rho(770)$ and the
$\phi(1020)$ mesons. This property is also verify for each orbital excitation
of these states. This behavior is well reproduced within the usual potential
models. But the strong relation between orbital and radial trajectories
obtained for the energy in the previous Sections shows that this remarkable
property will be also verify for radial excitations within a potential model
description. This leads to a mass for the $K^*(2S)$ of about 1565 MeV which is
approximately the value found with the usual models; this value is
just between the masses of the two possible candidates for this states
$K^*(1410)$ and $K^*(1680)$. This is a major problem of usual potential
models: the radial excitations of light strange mesons cannot be
satisfactorily be described (see for example~\cite{brau98,sema97,bura97}).

\section{The one-dimensional BSQ approach}
\label{sec:one}

In this section we show that a 1-dimensional BSQ approach can be used to
derive some simple formulae which can be applied to the 3-dimensional $\ell=0$
states. For a central potential, odd states of the 1-dimensional
Schr\"{o}dinger equation remain solutions of the 3-dimensional $\ell=0$
Schr\"{o}dinger equation (considering only the $x \geq 0$ part of the
$x$-axis). This property can be used within a BSQ approach.

In a 3-dimensional calculation, if we use the Langer prescription, the
centrifugal term never vanishes. But in a pure 1-dimensional case, the absence
of this centrifugal term simplifies the evaluation of the action variable and
leads to very simple formulae. Indeed, in the non-relativistic case, we have
\begin{equation}
\label{eq44}
p=\pm  \sqrt{2\mu(E-V(x))}.
\end{equation}
With $V(x)=a|x|^{\alpha}$, the quantization of the action variable $J$ leads
to
an equation which can be solved for the energy:
\begin{equation}
\label{eq45}
E =\left(\frac{a^2}{(2\mu)^{\alpha}}\right)^{1/(\alpha+2)} 
\left(\frac{\alpha \pi}{B(1/\alpha,3/2)}\ (n+3/4)\right)^{2\alpha/(\alpha+2)},
\end{equation}
where we have changed $n+1/2$ into $(2n+1)+1/2$ to only take into account odd
states. This last
result is an 
extension to the three-dimensional case of a formula obtained previously from
a one-dimensional calculation~\cite{cari93}. This formula is very close from
the Eq.~(\ref{eq12}), but here it can
be used, in principle, to approximate all the 3-dimensional $\ell = 0$ states
of a pure power-law potential.

\begin{figure}
\centering
\includegraphics*[width=12cm]{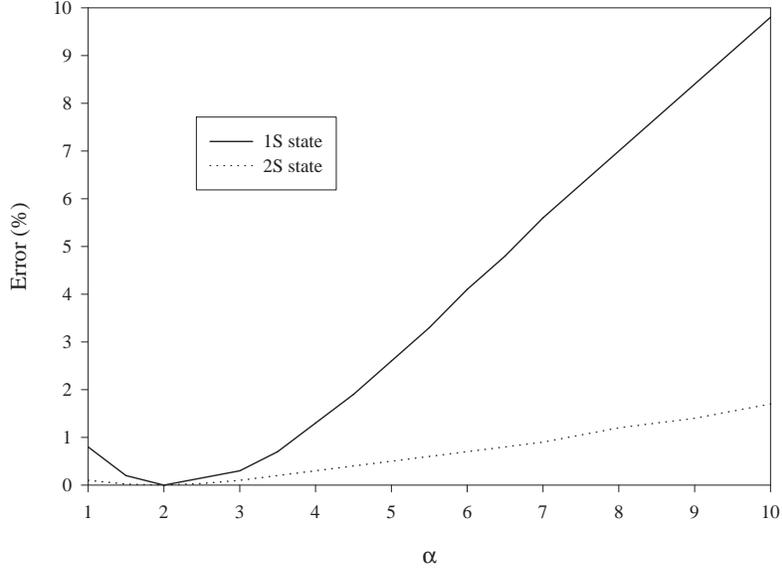}
\caption{Evolution of the errors introduced by Eq.~(\protect\ref{eq45}) as a function of $\alpha$ for the ground state and the first radial excitation.}
\label{fig:1}
\end{figure}

It is well known, from scaling properties (see for example~\cite{luch91}),
that the energy obtained with the Schr\"{o}dinger equation for a power-law
potential can be written as
\begin{equation}
\label{eq46}
E =\left(\frac{a^2}{(2\mu)^{\alpha}}\right)^{1/(\alpha+2)} \epsilon,
\end{equation}
where $\epsilon$ is a solution of a dimensionless Schr\"{o}dinger equation.
Since the BSQ approach gives the correct dimensional factor, the error depends
only on $\alpha$ and $n$. Fig.~\ref{fig:1} shows the evolution of errors with
$\alpha$ for $n=1$ and $n=2$. For the ground states the errors are about $1\%$
(or smaller) for $\alpha \leq 4$. As expected, the errors decrease rapidly
with the increase of the radial quantum number.

In general it is not obvious to understand why the formula (\ref{eq45}) works
so well since the exact solution is not known. For the harmonic oscillator
case this formula gives the correct position of the energy levels. But it is
more instructive to consider the case of a linear potential. Indeed, for
$\alpha=1$ Eq.~(\ref{eq45}) leads to 
\begin{equation}
\label{eq47}
\epsilon =\left(\frac{3\pi}{4}(n+3/4)\right)^{2/3}\equiv z^{2/3}.
\end{equation}
This expression is just the leading term of an expansion ($z \gg 1$) for the
values of the zeros of the Airy function~\cite [p. 450] {abra70}. The exact
solution is obtained by the summation of all the terms present in the
expansion. In this particular case we can see how the Eq.~(\ref{eq45})
approximates the exact solution and how evolves this approximation. The
formula (\ref{eq47}) gives much better results than variational methods using
trial wave functions for which the accuracy decrease rapidly as a function of
$n$ (see for example~\cite [p. 267] {flam86}).

It is worth noting that a 3-dimensional calculation using BSQ approach
introduces smaller errors than the 1-dimensional formulae obtained here. For
example, for a linear potential the error is about $0.5\%$ for the ground
state when a 3-dimensional calculation is performed, and about $0.75\%$ with
the 1-dimensional formula. But in general the 3-dimensional calculations do
not lead to analytical formulae or lead to complicated ones.

We can also calculate the mean square radius for a power-law potential. Using
the definition (\ref{eq17}) we can write:
\begin{equation}
\label{eq48}
\langle r^2 \rangle = \left(\frac{1}{2 a \mu}\right)^{2/(\alpha+2)} \,
\frac{B(3/\alpha,1/2)}{B(1/\alpha,1/2)}\,
\left(\frac{\alpha \pi}{B(1/\alpha,3/2)}\ (n+3/4)\right)^{4/(\alpha+2)}.
\end{equation}
The calculation of the values of the wave functions at the origin leads to
\begin{equation}
\label{eq49}
|\Psi(0)|^2= \frac{\mu\alpha\, a^{1/\alpha}}{\pi B(1/\alpha,1/2)}\
E^{(\alpha-1)/\alpha}, 
\end{equation}
but this time with the expression (\ref{eq45}) for the energy.
The evolution of the accuracy of formulae (\ref{eq48}) and (\ref{eq49})
with $\alpha$ and $n$ is similar to that shown in Fig.~\ref{fig:1} for the
energy.

The formula for the ratio of the wave function at the origin is given by
\begin{equation}
\label{eq50}
\left|\frac{\Psi_m(0)}{\Psi_n(0)}\right|^2=\left(\frac{m+3/4}{n+3/4}\right)^{2
(\alpha-
1)/(\alpha+2)}.
\end{equation}
Some connection with previous general results can be done.
If we suppose that $m=2$ and $n=1$, formula (\ref{eq50}) shows that the
ratio is greater than 1 if $\alpha > 1$ $(d^2V(r)/dr^2 >0)$ and smaller than 1
if $\alpha <1$ $(d^2V(r)/dr^2 <0)$. This behavior is predicted by a general
result obtained in Ref.~\cite{mart77}. Eq.~(\ref{eq45}) shows that the energy
behaves with the reduced mass as $E \propto \mu^{-\alpha/(\alpha+2)}$. Thus we
can
calculate that  
\begin{equation}
\label{eq50b}
\frac{d}{d\mu}\left(\frac{|\Psi_n(0)|^2}{\mu}\right)\propto
\frac{1-\alpha}{\alpha+2}
\, \mu^{-(1+2\alpha)/(\alpha+2)}.
\end{equation}
This quantity is positive if $\alpha <1$ $(d^2V(r)/dr^2 <0$ and $dV(r)/dr\geq
0)$ and negative if $\alpha>1$. This property is proved for $n=1$ in
Ref.~\cite{rosn78}.

The same calculations, for the energy and the r.m.s. radius, can be done for a
semirelativistic kinematics but unfortunately they lead to complicated
equations involving hypergeometric functions. The relativistic version of
Eq.~(\ref{eq45}) can only be solved explicitly for the energy when one
considers asymptotic behaviors. Thus further information about the observables
cannot be extracted within the frame of a 1-dimensional semirelativistic
approach.

To conclude this Section, we show that the 1-dimensional BSQ approach can
also be used to derive a simple formula giving the number of $\ell = 0$ bound
states (a generalization to $\ell \not= 0$ states can easily be done). We
consider an attractive potential which vanishes at infinity (for
example a Gaussian or a Yukawa potential). We calculate the value of
the radial quantum number $n$ for $E=0$:
\begin{equation}
\label{eq51}
n=\frac{2\sqrt{2\mu}}{\pi}\int_0^{\infty} \sqrt{-V(r)}\, dr-\frac{1}{2}.
\end{equation}
In general this number is not an integer except if the energy
level $E=0$ is a real solution. But the integer part of $n$ gives the radial
quantum number $n_{max}$ of the highest energy level. The formula for the
number
of bound states, $N=n_{max}+1$, of a given potential reads
\begin{equation}
\label{eq52}
N=\left[\frac{2\sqrt{2\mu}}{\pi}\int_0^{\infty} \sqrt{-V(r)}\,
dr+\frac{1}{2}\right],
\end{equation}
where $[x]$ denotes the integer part of $x$. For example, if $V(r)$ is of the
form $V(r)=-a f(br)$, we have
\begin{equation}
\label{eq53}
N=\left[\frac{2\sqrt{2\mu a}}{\pi b}\int_0^{\infty} \sqrt{f(y)}\,
dy+\frac{1}{2}\right].
\end{equation}
The remaining integral is a pure number and we can see immediately the
dependence of the number of bound states on the potential parameters.

The same calculation can be performed in the semirelativistic case and we find
\begin{equation}
\label{eq54}
N=\left[\frac{1}{\pi}\int_0^{\infty} \sqrt{V(r)^2-4 m V(r)}\,
dr+\frac{1}{2}\right].
\end{equation}
In this case the simple reduction performed in Eq.~(\ref{eq53}) cannot be
obtained.

\section{Summary}
\label{sec:summary}

We have shown in this paper that the Bohr-Sommerfeld quantization procedure is
an accurate and powerful method. It makes possible the derivation of reliable
analytical formulae, from which one can easily study, e.g., the dependence of
some physical quantities on the quantum numbers. Our study emphasizes the
strong connection existing between the Regge and radial trajectories for the
energy within a potential model, which could play an important role in 
elucidating the confining properties of the quark-antiquark interaction when
additional experimental information is obtained on the radial excitations of
light mesons; the study of other observables (decay widths, mean square radii,
electromagnetic mass splittings,...) could put supplementary constraints on
this interaction. 

We have also shown that even if semirelativistic calculations yield results
quantitatively different compared with nonrelativistic calculations some
common general features of observables are observed. The asymptotic behavior
of orbital and radial trajectories, for the energy and the r.m.s. radius, are
the same for a potential which behaves asymptotically as a power-law. Moreover
the ratio of the slopes of these trajectories depends only on the value of the
power of the confining potential. 

At last, we have shown in Sec.~\ref{subsec:thirdf} that 1-dimensional
calculations lead to very simple approximated formula for the energy, the
r.m.s. radius and the wave function at the origin in the $\ell = 0$ sector. We
have also obtained formulae which give the number of bound states of a given
potential for the nonrelativistic and semirelativistic case.

Of course, a BSQ approach cannot replace a correct quantum
description since it is only an approximate method which is not completely
self-consistent. For example, we need to use the Schr\"{o}dinger equation to
give a definition of the wave function at the origin to be able to calculate
this quantity with this formalism. Moreover some quantum concepts have no
meaning in this framework. But we emphasize that for some cases the evaluation
of observables (evaluation of average values) can be performed using this old
method; even if calculations yield complicated formulae it is often possible
to extract interesting information from an analytical relation and in this way
guide full quantum calculations.

\acknowledgments

We thank Professor F. Michel and Dr. C. Semay for their availability and
useful discussions. We also would like to thank IISN for financial support.

\clearpage
\begin{table}
\squeezetable
\protect\caption{Energies, r.m.s. radii and wave functions at the origin for a
selected set of mesons, obtained from the Schr\"{o}dinger equation and from
the BSQ approximation for the Cornell potential, using the parameters of
Ref.~\protect\cite{fulc94} (masses are given in MeV, r.m.s. radii in
GeV$^{-1}$, and wave functions at the origin in GeV$^{3}$).}
\label{tab:1}
\begin{tabular}{ccccccc}
States & Exact Masses & BSQ & Exact $\sqrt{\bar{r}^2}$ & BSQ &
Exact $|\Psi(0)|^2$& BSQ \\  
\hline
$n\bar{n}$ states &  &  &  &  &  &\\
$1S$ & 681  & 681  & 4.04 & 3.97 & 7.5715 10$^{-3}$& 7.6383 10$^{-3}$ \\
$2S$ & 1577 & 1577 & 7.32 & 7.30 & 6.7084 10$^{-3}$ & 6.7601 10$^{-3}$\\
$1P$ & 1240 & 1242 & 5.82 & 5.79 & & \\
$1D$ & 1692 & 1693 & 7.32 & 7.31 & &\\
$n\bar{s}$ states &  &  &  &  \\
$1S$ & 794  & 794  & 3.65 & 3.58 & 1.0653 10$^{-2}$& 1.0784 10$^{-2}$ \\
$2S$ & 1626 & 1626 & 6.68 & 6.65 & 9.2169 10$^{-3}$& 9.3099 10$^{-3}$\\
$1P$ & 1320 & 1321 & 5.30 & 5.28 & & \\
$1D$ & 1738 & 1739 & 6.69 & 6.67 & & \\
$s\bar{s}$ states &  &  &  &  \\
$1S$ & 1004 & 1002 & 3.17 & 3.10 & 1.7346 10$^{-2}$& 1.7668 10$^{-2}$ \\
$2S$ & 1759 & 1758 & 5.87 & 5.85 & 1.4416 10$^{-2}$& 1.4623 10$^{-2}$\\
$1P$ & 1490 & 1492 & 4.67 & 4.65 & & \\
$1D$ & 1869 & 1870 & 5.91 & 5.90 & & \\
$n\bar{c}$ states &  &  &  &  \\
$1S$ & 1973 & 1972 & 3.36 & 3.29 & 1.4194 10$^{-2}$& 1.4418 10$^{-2}$\\
$2S$ & 2757 & 2757 & 6.19 & 6.17 & 1.2004 10$^{-2}$& 1.2154 10$^{-2}$\\
$1P$ & 2474 & 2475 & 4.92 & 4.89 & & \\
$2P$ & 3125 & 3126 & 7.43 & 7.42 & & \\
$c\bar{c}$ states &  &  &  &  \\
$1S$ & 3067 & 3062 & 2.26 & 2.19 & 5.8025 10$^{-2}$& 6.0210 10$^{-2}$ \\
$2S$ & 3693 & 3691 & 4.38 & 4.36 & 4.2153 10$^{-2}$& 4.3351 10$^{-2}$\\
$1P$ & 3497 & 3497 & 3.48 & 3.46 & & \\
$2P$ & 3991 & 3991 & 5.35 & 5.34 & & \\
$1D$ & 3806 & 3806 & 4.47 & 4.46 & & \\
$2D$ & 4242 & 4242 & 6.17 & 6.17 & & \\
$n\bar{b}$ states &  &  &  &  \\
$1S$ & 5313 & 5311 & 3.16 & 3.09 & 1.7607 10$^{-2}$& 1.7937 10$^{-2}$\\
$2S$ & 6066 & 6065 & 5.85 & 5.83 & 1.4613 10$^{-2}$& 1.4825 10$^{-2}$\\
$1P$ & 5799 & 5800 & 4.65 & 4.63 & & \\
$2P$ & 6420 & 6420 & 7.04 & 7.03 & & \\
$b\bar{b}$ states &  &  &  &  \\
$1S$ & 9448  & 9439  & 1.13 & 1.04 & 7.3583 10$^{-1}$& 7.7015 10$^{-1}$\\
$2S$ & 10007 & 10003 & 2.55 & 2.52 & 3.3903 10$^{-1}$& 3.5597 10$^{-1}$\\
$1P$ & 9901  & 9900  & 2.04 & 2.02 & & \\
$2P$ & 10261 & 10261 & 3.28 & 3.28 & & \\
$1D$ & 10148 & 10148 & 2.74 & 2.73 & & \\  
\end{tabular}
\end{table}

\begin{table}
\squeezetable
\protect\caption{Energies and r.m.s. radii for a selected set of mesons,
obtained from
the spinless Salpeter equation and from the BSQ approximation for the Cornell
potential, 
using the parameters of Ref.~\protect\cite{fulc94} (masses are given 
in MeV and r.m.s. radii in GeV$^{-1}$).}
\label{tab:2}
\begin{tabular}{ccccc}
States & Exact Masses & BSQ & Exact $\sqrt{\bar{r}^2}$ & BSQ \\  
\hline
$n\bar{n}$ states &  &  &  &  \\
$1S$ & 703  & 683  & 3.30 & 3.22  \\
$2S$ & 1416 & 1413 & 5.40 & 5.39  \\
$1P$ & 1240 & 1236 & 4.62 & 4.60  \\
$1D$ & 1642 & 1639 & 5.60 & 5.60  \\
$s\bar{s}$ states &  &  &  &  \\
$1S$ & 1004 & 991  & 2.96 & 2.87  \\
$2S$ & 1695 & 1691 & 5.04 & 5.02  \\
$1P$ & 1508 & 1506 & 4.27 & 4.25  \\
$1D$ & 1885 & 1884 & 5.27 & 5.26  \\
$c\bar{c}$ states &  &  &  &  \\
$1S$ & 3067 & 3056 & 2.05 & 1.97  \\
$2S$ & 3668 & 3662 & 3.89 & 3.85  \\
$1P$ & 3504 & 3504 & 3.19 & 3.18  \\
$2P$ & 3970 & 3970 & 4.76 & 4.76  \\
$1D$ & 3811 & 3811 & 4.07 & 4.06  \\
$2D$ & 4216 & 4216 & 5.48 & 5.48  \\
$b\bar{b}$ states &  &  &  &  \\
$1S$ & 9448  & 9433  & 1.10 & 1.01  \\
$2S$ & 9999  & 9993  & 2.45 & 2.42  \\
$1P$ & 9900  & 9900  & 1.99 & 1.98  \\
$2P$ & 10262 & 10262 & 3.17 & 3.17  \\
$1D$ & 10150 & 10150 & 2.66 & 2.66  \\  
\end{tabular}
\end{table}

\end{document}